\documentclass[a4paper,11pt]{article}
\usepackage{jheppub}
\usepackage[T1]{fontenc}

\newcommand{\BE}{\begin{equation}}
\newcommand{\EE}{\end{equation}}
\newcommand{\BA}{\begin{eqnarray}}
\newcommand{\EA}{\end{eqnarray}}

\title{\boldmath Brane-Antibrane at Finite Temperature in the Framework of Thermo Field Dynamics}

\author[a]{Kenji Hotta,}

\affiliation[a]{Department of Physics, Hokkaido University, Sapporo, Hokkaido, 060-0810, Japan}

\emailAdd{khotta@particle.sci.hokudai.ac.jp}

\abstract{Previously we have investigated the thermodynamical properties of D-brane--anti-D-brane pairs by calculating the one-loop free energy and the finite temperature effective action in the framework of Matsubara formalism. This calculation is based on boundary string field theory, and we have a problem of a choice of Weyl factors on the two boundaries of one-loop open string world-sheet. We have chosen one of them and compute the one-loop free energy. In this paper, we derive the mass spectrum of open strings on a D-brane--anti-D-brane pair supposing that the world-sheet is a strip, and compute the thermal vacuum state, free energy and the partition function for a single open string in the framework of thermo field dynamics. From this we can reproduce the free energy for multiple strings in the case of Matsubara method. This fact reinforces our argument of the choice of the Weyl factors in the case of Matsubara formalism.}

\begin{document} 
\maketitle
\flushbottom

\section{Introduction}
\label{sec:Intro}

Since the late 1980s, a considerable number of studies have been made on thermodynamical behavior of strings continuously. Salient feature of finite temperature system of strings appears near the Hagedorn temperature, which is the upper limit temperature for perturbative strings. There is a possibility that enigmas related to black holes and the early universe are resolved by pursuing thermodynamics of strings and branes. The finite temperature systems of strings and branes have been mainly investigated in the framework of Matsubara formalism \cite{Matsubara}. In this formalism, we can obtain partition function or free energy by computing amplitudes in the space where the imaginary time direction is compactified with period $\beta$. In the case of ideal gas of strings, free energy can be approximated by one-loop amplitude, and it diverges above the Hagedorn temperature.

We have previously discussed the behavior of coincident D-brane--anti-D-brane pairs at finite temperature in the framework of Matsubara formalism \cite{Hotta4} \cite{Hotta5} \cite{Hotta6}. We have calculated the finite temperature effective potential of open strings on these branes based on boundary string field theory (BSFT). Although D-brane--anti-D-brane pairs are unstable at zero temperature, there are the cases that they become stable at finite temperature. Let us consider the 10-dimensional flat spacetime case as an example. For the D9-$\overline{\textrm{D9}}$ pairs, a phase transition occurs at slightly below the Hagedorn temperature and the D9-$\overline{\textrm{D9}}$ pairs become stable above this temperature. On the other hand, for the D$p$-$\overline{\textrm{D$p$}}$ pairs with $p \leq 8$, such a phase transition does not occur. We thus concluded that not a lower dimensional brane-antibrane pairs but D9-$\overline{\textrm{D9}}$ pairs are created near the Hagedorn temperature in this case. Let us call this phase transition `{\it thermal brane creation transition}'. These works are generalized by Cal\`{o} and Thomas to the case that a D-brane and an anti-D-brane are separated \cite{Separated}. We have applied these works to cosmology \cite{Hotta7}. Then we have proposed a conjecture that {\it D9-brane--$\overline{\textrm{D9}}$-brane pairs are created by the Hagedorn transition of closed strings,} and describe some circumstantial evidences for this conjecture \cite{Hotta8}.

Unfortunately, there is a problem in the calculation of one-loop amplitude based on BSFT \cite{1loopAO} \cite{1loopann1} \cite{1loopann2} \cite{1loopsym1} \cite{1loopsym2} \cite{1loop1} \cite{1loop2} \cite{1loop3} \cite{1loop4}. Since boundary action for open string world-sheet breaks the conformal invariance, there is no specific choice of Weyl factors in the two boundaries of the one-loop world-sheet. Our calculation is based on the one-loop amplitude of open strings, which has been proposed by Andreev and Oft \cite{1loopAO}. Their choice of Weyl factors is natural in the sense that both sides of the cylinder world-sheet are treated on an equal footing. Moreover, the low energy part of the one-loop amplitude of open string with this boundary action coincide with that of the tachyon field model \cite{tachyon1} \cite{tachyon2}. However, these facts do not fully guarantee the validity of this choice. In any case, if we can reproduce the same result by using another method, it reinforces our previous studies.

The finite temperature system of strings is mainly analyzed by using Matsubara method. We have calculated the one-loop free energy of open strings on D$p$-$\overline{\textrm{D$p$}}$ pairs following the recipe for this method. However, there are other methods to investigate finite temperature system. One of the useful methods is thermo field dynamics (TFD), which was first proposed by Takahashi and Umezawa \cite{TFD}. In this method, we introduce fictitious copy of system, and consider doubled system. It is expected that this method is available to non-equilibrium systems, since TFD is a real time formalism unlike Matsubara formalism.

Quite a few works have been done to investigate finite temperature systems of strings in the framework of TFD, instead of Matsubara formalism. TFD has already been applied to second quantized string field theory in 1980s \cite{SFTTFD1} \cite{SFTTFD2} \cite{SFTTFD3}, and few years ago \cite{SFTTFD4}. However, recently these systems have been investigated mainly based on first quantized string theory, since string field theory is quite difficult in general. For example, it is applied to the D-brane case \cite{DbraneTFD1} \cite{DbraneTFD2} \cite{DbraneTFD3} \cite{DbraneTFD4} \cite{DbraneTFD5} and the closed string case \cite{ClosedStringTFD}.

The purpose of this paper is to compute free energy and partition function for a single open string on a D$p$-$\overline{\textrm{D$p$}}$ pair in the framework of TFD, and then derive free energy for multiple open strings on these branes. Consequently, we can reproduce the one-loop free energy for open strings on these branes in the framework of Matsubara formalism. It is important to note that we need not handle the one-loop amplitude of open strings in this case. This result leads to the conclusion that our choice of Weyl factors in the case of Matsubara method is reasonable.

The outline of this paper is as follows. We first review the calculation in the framework of Matsubara formalism in \S 2. We describe the partition function in terms of light-cone momentum, and the mass spectrum of open strings on a D$p$-$\overline{\textrm{D$p$}}$ pair when the world-sheet is a strip in \S 3. In \S 4 and \S 5, we compute the thermal vacuum states, free energy and partition function for a single string in the framework of TFD. We derive free energy for multiple strings from the partition function for a single string in \S 6. Finally, \S 7 presents our conclusions and directions for future work.

\section{Brane-antibrane at Finite Temperature in the Framework of Matsubara Formalism}
\label{sec:Matsubara}

Let us begin by reviewing the calculation of the one-loop free energy and the finite temperature effective potential of open strings on a D$p$-$\overline{\textrm{D$p$}}$ pair in the framework of Matsubara formalism \cite{Hotta4} \cite{Hotta5} \cite{Hotta6}. A D$p$-$\overline{\textrm{D$p$}}$ pair is unstable at zero temperature, and open strings on them has tachyonic mode corresponding to the lowest energy state. This open string tachyon can be treated by BSFT. In this theory we introduce 1-dimensional boundary action in addition to 2-dimensional world-sheet bulk action. If we consider the constant tachyon case, the boundary action is given by
\BE
  I_{\partial \Sigma}
    = \int_{\partial \Sigma} d \sigma^a \ \sqrt{- h_{aa}} \ |T|^2,
\label{eq:DiskBoundaryAction}
\EE
where $T$ is a complex scalar tachyon field on a D$p$-$\overline{\textrm{D$p$}}$ pair. $\sigma^a$ parametrizes the boundary of the world-sheet $\Sigma$, and $h_{aa}$ is the induced metric on the boundary. The potential of this complex tachyon field $T$ can be derived based on BSFT as \cite{tachyon1} \cite{tachyon2} \cite{TakaTeraUe}
\BE
  V(T) = 2 \tau_p v_p \exp (-8 |T|^2),
\EE
by computing tree level amplitude. $v_p$ is the $p$-dimensional volume of the system that we are considering. This potential has the maximum at $T=0$, and the minimum at $|T| = \infty$ . We can adjust the height of the potential, such that it coincides with the tension energy of the branes. Then $\tau_p$ represents the tension of a D$p$-brane, which is defined by
\BE
  \tau_p = \frac{1}{(2 \pi)^p
    {\alpha'}^{\scriptscriptstyle \frac{p+1}{2}} g_s},
\label{eq:tension}
\EE
where $g_s$ is the coupling constant of strings and $\alpha'$ is the slope parameter, which is the only dimensionful parameter string theory has. This potential satisfies Sen's conjecture, that is, tachyon condensation on brane-antibrane is equivalent to the disappearance of them \cite{Senconjecture1} \cite{Senconjecture2} \cite{Senconjecture3} \cite{Senconjecture4}.

We have calculated one-loop free energy of open strings on these branes in the constant tachyon background based on boundary string field theory. In this case we are confronted with the problem of the choice of Weyl factors in the two boundaries of the one-loop world-sheet, because the conformal invariance is broken by boundary action. The calculation in the previous work is based on the one-loop amplitude of open strings, which has been proposed by Andreev and Oft \cite{1loopAO}. In this case, the boundary action of string world-sheet is given by
\BE
  I_{\partial \Sigma}
    = \int_{0}^{2 \pi \tau} d \sigma_0 \int_{0}^{\pi} d \sigma_1
      \left\{ |T|^2 \delta (\sigma_1)
        + |T|^2 \delta (\pi - \sigma_1) \right\}.
\label{eq:CylinderBoundaryAction}
\EE
Their choice of Weyl factors is natural in the sense that both sides of the cylinder world-sheet are treated on an equal footing. Moreover, the low energy part of the one-loop amplitude of open string with this boundary action coincide with that of the tachyon field model \cite{tachyon1} \cite{tachyon2}. Using this boundary action, we can obtain the one-loop free energy as
\BA
  F (T, \beta)
    &=& - \ \frac{2 (2 \pi)^4 v_p}{\beta_H^{p+1}}
      \int_{0}^{\infty} \frac{d \tau}{\tau} \ 
        \tau^{- \frac{p+1}{2}} e^{-4 \pi |T|^2 \tau}
          \nonumber \\
  && \hspace{20mm}
    \times \left[ \left\{ \frac{\vartheta_3 (0 | i \tau)}
      {{\vartheta_1}' (0 | i \tau)} \right\}^4
        \left\{ \sum_{w=1}^{\infty}
          \exp \left( - \frac{\pi w^2 \beta^2}{\beta_H^2 \tau}
            \right) \right\}
              \right. \nonumber \\
  && \hspace{30mm} \left.
    - \left\{ \frac{\vartheta_2 (0 | i \tau)}
      {{\vartheta_1}' (0 | i \tau)} \right\}^4
        \left\{ \sum_{w=1}^{\infty} (-1)^w
          \exp \left( - \frac{\pi w^2 \beta^2}{\beta_H^2 \tau}
            \right) \right\} \right].
\label{eq:Dp_antiDp_Free}
\EA
It should be noted that we can obtain the free energy for multiple strings by only calculating one-loop amplitude in this background, because the world-sheet multiple winding around the compactified Euclidean time represents multiple strings. The finite temperature effective potential can be obtained as a sum of the tachyon potential at zero temperature and free energy of open strings. We can discuss the thermodynamical behavior of D$p$-$\overline{\textrm{D$p$}}$ pairs from the potential minimum of this finite temperature effective potential.

The results are summarized as follows. For the D9-$\overline{\textrm{D9}}$ pairs, a phase transition occurs at slightly below the Hagedorn temperature and the D9-$\overline{\textrm{D9}}$ pairs become stable above this temperature. On the other hand, for the D$p$-$\overline{\textrm{D$p$}}$ pairs with $p \leq 8$, such a phase transition does not occur. We thus concluded that not a lower dimensional brane-antibrane pairs but D9-$\overline{\textrm{D9}}$ pairs are created near the Hagedorn temperature. We have also investigated the non-BPS D-brane case and concluded that non-BPS D9-branes are created near the Hagedorn temperature \cite{Hotta6}. The spacetime-filling branes become stable near the Hagedorn temperature, not only in a flat spacetime but also in a toroidally compactified spacetime \cite{Hotta5} \cite{Hotta6}.

\section{Light-cone Momentum and Mass Spectrum}
\label{sec:LCMass}

Before moving to the calculation based on TFD, we rewrite the partition function using light-cone momentum for the later convenience, and describe the mass spectrum of open strings on a D$p$-$\overline{\textrm{D$p$}}$ pair. It is favorable to argue this system based on string field theory. However, we treat an open string on these branes as a first quantized string, since second quantized string field theory is quite difficult. The partition function for a single string is given by
\BA
  Z_1 (\beta)
    = \textrm{Tr} \ \exp \left( - \beta k^0 \right),
\EA
where $k^0$ is energy for a single string. It is convenient to introduce $k^+$ and $k^-$ as
\BA
  k^+ &=& \frac{1}{\sqrt{2}} \left( k^0 + k^1 \right),
\label{eq:p+} \\
  k^- &=& \frac{1}{\sqrt{2}} \left( k^0 - k^1 \right),
\EA
and to rewrite this partition function as
\BA
  Z_1 (\beta)
    = \textrm{Tr} \ 
      \exp \left[ - \ \frac{1}{\sqrt{2}} \ \beta \left( k^+
        + \frac{| \mbox{\boldmath $k$} |^2 + M^2}{2 k^+} \right) \right],
\label{eq:Z1LC}
\EA
where we have used the dispersion relation
\BA
  2 k^+ k^- - | \mbox{\boldmath $k$} |^2 - M^2 = 0.
\EA
Substituting mass spectrum of open strings on a D$p$-$\overline{\textrm{D$p$}}$ pair into (\ref{eq:Z1LC}), we obtain the partition for a single string $Z_1 (\beta)$.

Let us derive the mass spectrum of open superstrings on a D$p$-$\overline{\textrm{D$p$}}$ pair. In order to get the finite temperature effective potential, we need to know the mass of open strings in a constant tachyon background. We suppose that the world-sheet has the form of a strip and that the range of the spatial coordinate $\sigma^1$ is $0 \leq \sigma^1 \leq \pi$. The boundary of the world-sheet is parametrized by the time coordinate $\sigma^0$. The world-sheet action in the flat Minkovski space-time has two parts, the bulk part and the boundary one as
\BA
  I_2 = I_{\Sigma} + I_{\partial \Sigma},
\EA
where
\BA
  I_{\Sigma} &=& - \ \frac{1}{4 \pi \alpha'} \int d^2 \sigma \ 
    \sqrt{-g} \ g^{\alpha \beta} \eta_{\mu \nu}
      \left( \partial_{\alpha} X^{\mu} \partial_{\beta} X^{\nu}
        - i \alpha' {\overline{\psi}}^{\mu} \rho_{\alpha}
          \partial_{\beta} \psi^{\nu} \right), \\
  I_{\partial \Sigma}
    &=& \int_{\Sigma} d^2 \sigma \sqrt{- h_{00}} \ |T|^2
      \left\{ \delta (\sigma^1) + \delta (\pi - \sigma^1) \right\}.
\EA
We have denoted two-dimensional Dirac matrices as $\rho_{\alpha}$. The induced metric $h_{00}$ is calculated as
\BA
  h_{00} = g_{\alpha \beta} \ 
    \frac{\partial \sigma^{\alpha}}{\partial \sigma^0} \ 
      \frac{\partial \sigma^{\beta}}{\partial \sigma^0}
        = g_{00}.
\EA
We have assumed that the boundary action has the same form on both edges. From this action, the world-sheet energy-momentum tensor can be calculated via the variation of the action with respect to the metric
\BA
  T_{\alpha \beta}
    = - 4 \pi \alpha' \ \frac{1}{\sqrt{-g}} \ 
      \frac{\delta I_2}{\delta g^{\alpha \beta}}.
\EA
The mass spectrum of open strings is derived from the Virasoro constraint
\BA
  T_{\alpha \beta} = 0.
\EA
The bulk part of the energy-momentum tensor can be calculated as usual. The components of energy-momentum tensor are given by
\BA
  T_{00}^{\Sigma} &=& T_{11}^{\Sigma}
    = \frac{1}{2} \ \eta_{\mu \nu}
      \left( \partial_0 X^{\mu} \partial_0 X^{\nu}
        + \partial_1 X^{\mu} \partial_1 X^{\nu} \right)
          + \frac{i}{2} \ \alpha' \eta_{\mu \nu}
            \left( \psi_-^{\mu} \partial_0 \psi_-^{\nu}
              + \psi_+^{\mu} \partial_0 \psi_+^{\nu}
                \right), \\
  T_{01}^{\Sigma} &=& T_{10}^{\Sigma}
    = \eta_{\mu \nu}
      \partial_0 X^{\mu} \partial_1 X^{\nu}
        - \ \frac{i}{4} \ \alpha' \eta_{\mu \nu}
          \left\{ \psi_-^{\mu}
            \left( \partial_0 - \partial_1 \right) \psi_-^{\nu}
              - \psi_+^{\mu}
                \left( \partial_0 + \partial_1 \right) \psi_+^{\nu}
                  \right\},
\EA
where we have set the world-sheet metric as
\BA
  g_{\alpha \beta} =
    \left(
      \begin{array}{cc}
        -1 & 0 \\
        0 & 1 \\
      \end{array}
    \right),
\EA
and chosen the basis of Dirac matrices as
\BA
  \rho^0 =
    \left(
      \begin{array}{cc}
        0 & -i \\
        i & 0 \\
      \end{array}
    \right), \ \ \ 
  \rho^1 =
    \left(
      \begin{array}{cc}
        0 & i \\
        i & 0 \\
      \end{array}
    \right).
\EA
We have represented the components of $\psi^{\mu}$ in this basis as
\BA
  \psi^{\mu} =
    \left(
      \begin{array}{c}
        \psi_-^{\mu} \\
        \psi_+^{\mu} \\
      \end{array}
    \right).
\EA
Let us explain the derivation of the boundary part of the energy-momentum tensor. Since $g^{\alpha \beta}$ is the inverse matrix of $g_{\alpha \beta}$, namely,
\BA
  g_{\alpha \beta} g^{\beta \gamma} = \delta_{\alpha}^{\gamma},
\EA
$h_{00} = g_{00}$ can be represented as
\BA
  h_{00} = g g^{11}.
\EA
From this we can calculate the variation of $\sqrt{- h_{00}}$ with respect to the metric as
\BA
  \frac{\delta \sqrt{- h_{00} (\sigma')}}
    {\delta g^{\alpha \beta} (\sigma)}
      = \left\{ - \frac{1}{2} \sqrt{- h_{00} ({\sigma^0} ')} \ 
        g_{\alpha \beta} (\sigma ')
          - \frac{1}{2 \sqrt{- h_{00} (\sigma')}} \ 
            g (\sigma ') \delta_{\alpha}^1 \delta_{\beta}^1 \right\}
              \delta^2 (\sigma - \sigma').
\EA
The energy-momentum tensor from the boundary action is given by
\BA
  T_{\alpha \beta}^{\partial \Sigma}
    = - 2 \pi \alpha' |T|^2 \left( - g_{\alpha \beta}
      + \delta_{\alpha}^1 \delta_{\beta}^1 \right)
        \left\{ \delta (\sigma^1) + \delta (\pi - \sigma^1) \right\}.
\EA
The components of the energy-momentum tensor are
\BA
  T_{00}^{\partial \Sigma}
    &=& - 2 \pi \alpha' |T|^2
      \left\{ \delta (\sigma^1) + \delta (\pi - \sigma^1) \right\}, \\
  T_{01}^{\partial \Sigma} &=& T_{10}^{\partial \Sigma}
    = T_{11}^{\partial \Sigma} = 0.
\EA
The total energy-momentum tensor $T_{\alpha \beta}$ is the sum of $T_{\alpha \beta}^{\Sigma}$ and $ T_{\alpha \beta}^{\partial \Sigma}$. From these results we can see that $T_{00}^{\partial \Sigma} \neq T_{11}^{\partial \Sigma}$ while $T_{00}^{\Sigma} = T_{11}^{\Sigma}$. At the boundary we cannot satisfy $T_{00} = 0$ and $T_{11} = 0$, simultaneously. We only keep the constraint $T_{00} = 0$ and omit $T_{11} = 0$ at the boundary. If we impose $T_{11} = 0$ instead of $T_{00} = 0$ then non-zero $T$ does not shift the mass of strings. However, it is unreasonable because the mass depends on the value of $T$ in the presence of tachyon potential $V (T)$.

Let us introduce the mode expansion of $X^{\mu}$ and $\psi^{\mu}$. In the constant tachyon case, the boundary action does not affect the boundary conditions of $X^{\mu}$ and $\psi^{\mu}$. These fields can have the Neumann boundary condition or the Dirichlet one. For simplicity, let us consider the all Neumann case, namely, the D9-$\overline{\textrm{D9}}$ pair case. In the lower dimensional brane case, we can similarly compute the mass spectrum. The mode expansion of $X^{\mu}$ is given by
\BA
  X^{\mu} (\sigma)
    = x^{\mu}
      + 2 \alpha' k^{\mu} \sigma^0
        + i \sqrt{\frac{\alpha'}{2}}
          \sum_{l \neq 0 , l = - \infty}^{\infty}
            \frac{1}{l} \ \alpha_l^{\mu}
              \left\{ \exp \left( - i l \sigma^- \right)
                + \exp \left( - i l \sigma^+ \right)
                  \right\},
\EA
and that of $\psi^{\mu}$ is given by
\BA
  \psi_-^{\mu}
    &=& \sum_{r \in Z + \frac{1}{2}}
      b_r^{\mu} \exp \left( - ir \sigma^- \right), \\
  \psi_+^{\mu}
    &=& \sum_{r \in Z + \frac{1}{2}}
      b_r^{\mu} \exp \left( - ir \sigma^+ \right),
\EA
for Neveu-Schwarz (NS) boundary condition, and
\BA
  \psi_-^{\mu}
    &=& \sum_{m \in Z}
      d_m^{\mu} \exp \left( - im \sigma^- \right), \\
  \psi_+^{\mu}
    &=& \sum_{m \in Z}
      d_m^{\mu} \exp \left( - im \sigma^+ \right),
\EA
for Ramond (R) boundary condition, respectively. We have defined $\sigma^{\pm} = \sigma^0 \pm \sigma^1$. The NS sector represents spacetime boson, and R one represents spacetime fermion. It is convenient to deal with the Virasoro constraint in the light-cone gauge
\BA
  X^+ (\sigma) &=& x^+ + 2 \alpha' k^+ \sigma^0, \\
  \psi^+ (\sigma) &=& 0,
\EA
where we have introduced the light-cone coordinates
\BA
  X^+ &=& \frac{1}{\sqrt{2}} \left( X^0 + X^1 \right), \\
  X^- &=& \frac{1}{\sqrt{2}} \left( X^0 - X^1 \right),
\EA
and their fermionic partners
\BA
  \psi^+ &=& \frac{1}{\sqrt{2}} \left( \psi^0 + \psi^1 \right), \\
  \psi^- &=& \frac{1}{\sqrt{2}} \left( \psi^0 - \psi^1 \right).
\EA
Substituting the mode expansions into the energy-momentum tensor and integrating over $\sigma^1$, we obtain
\BA
  k^- = \frac{1}{2 k^+} \left\{
    \sum_{I=2}^9 \left( k^I \right)^2
      + \frac{1}{\alpha'} \sum_{I=2}^9
        \left( \sum_{l=1}^{\infty} \alpha_{-l}^I \alpha_l^I
          + \sum_{r = \frac{1}{2}}^{\infty} r b_{-r}^I b_r^I \right)
            + 2 |T|^2 - \ \frac{1}{2} \right\},
\EA
in the NS sector, and
\BA
  k^- = \frac{1}{2 k^+} \left\{
    \sum_{I=2}^9 \left( k^I \right)^2
      + \frac{1}{\alpha'} \sum_{I=2}^9
        \left( \sum_{l=1}^{\infty} \alpha_{-l}^I \alpha_l^I
          + \sum_{m=1}^{\infty} m d_{-m}^I d_m^I \right)
            + 2 |T|^2 \right\},
\EA
in the R sector, respectively. We have presumed that the normal ordering constant is the same as usual if we take the limit as $T$ approaches zero. From these we can see that the mass square operators for NS string and R string are given by
\BA
  {M_{NS}}^2
    &=& \frac{1}{\alpha'}
      \left( N_B + N_{NS} + 2 |T|^2 - \ \frac{1}{2} \right),
\label{eq:MassNS} \\
  {M_R}^2
    &=& \frac{1}{\alpha'} \left( N_B + N_R + 2 |T|^2 \right),
\label{eq:MassR}
\EA
where the number operators for oscillation modes for boson, NS fermion and R fermion are defined as
\BA
  N_B &=& \sum_{l=1}^{\infty} \sum_{I=2}^9 \alpha_{-l}^I \alpha_l^I
    = \sum_{l=1}^{\infty} \alpha_{-l} \cdot \alpha_l, \\
  N_{NS} &=& \sum_{r = \frac{1}{2}}^{\infty} \sum_{I=2}^9 r b_{-r}^I b_r^I
    = \sum_{r = \frac{1}{2}}^{\infty} r b_{-r} \cdot b_r, \\
  N_R &=& \sum_{m=1}^{\infty} \sum_{I=2}^9 m d_{-m}^I d_m^I
    = \sum_{m=1}^{\infty} m d_{-m} \cdot d_m,
\EA
respectively. The dot denotes inner product on transverse space. $\alpha_l$ are operators for bosonic oscillators, and $b_r$ and $d_m$ are operators for NS fermionic oscillators and for R fermionic oscillators, respectively. The bosonic oscillators satisfy a commutation relation
\BA
  \left[ \alpha_m^I , \alpha_n^J \right]
    &=& m \delta_{m+n} \eta^{IJ},
\EA
and the fermionic oscillators anticommutation relations
\BA
  \left\{ b_r^I , b_s^J \right\} &=& \eta^{IJ} \delta_{r+s}, \\
  \left\{ d_m^I , d_n^J \right\} &=& \eta^{IJ} \delta_{m+n}.
\EA
From these mass spectra, we can see that a single string can be treated as a collection of bosonic and fermionic harmonic oscillators whose energy is shifted by constant tachyon and momentum of a string. We construct the thermal vacuum state for a single string in terms of these harmonic oscillators.

\section{Bogoliubov Transformations and Thermal Vacuum States}
\label{sec:BogVac}

Let us turn to the main subject. We here start to discuss the finite temperature system of a D$p$-$\overline{\textrm{D$p$}}$ pair in the framework of TFD, which has been proposed by Takahashi and Umezawa \cite{TFD}. In order to obtain the statistical average of a thermodynamical variable, we need to take trace of the corresponding operator. In TFD, it is replaced by the computation of the expectation value of the corresponding operator in a single state, namely, the thermal vacuum state. For this purpose, we introduce fictitious copy of the system, and duplicate the degrees of freedom of the system. In other words, we construct a copy of original Hilbert space, and prepare the tensor product of two copies of the Hilbert space. The thermal vacuum state is obtained from the Bogoliubov transformation of the vacuum state in this doubled Hilbert space. In this section, we describe this Bogoliubov transformation, and compute the thermal vacuum state.

As we have mentioned in the previous section, we can treat a single string as a collection of oscillators. In the case of a single open string on a D$p$-$\overline{\textrm{D$p$}}$ pair, the generators of the Bogoliubov transformation for NS string and R string are given by
\BA
  G_{1NS} &=& {\cal G}_B + {\cal G}_{NS}, \\
  G_{1R} &=& {\cal G}_B + {\cal G}_R,
\EA
where we have defined
\BA
  {\cal G}_B &\equiv& i \sum_{l=1}^{\infty} \frac{1}{l} \ \theta_l
    \left( \alpha_{-l} \cdot {\tilde{\alpha}}_{-l}
      - {\tilde{\alpha}}_l \cdot \alpha_l \right), \\
  {\cal G}_{NS} &\equiv& i \sum_{r = \frac{1}{2}}^{\infty} \ \theta_r
    \left( b_{-r} \cdot {\tilde{b}}_{-r}
      - {\tilde{b}}_r \cdot b_r \right), \\
  {\cal G}_R &\equiv& i \sum_{m=1}^{\infty} \ \theta_m
    \left( d_{-m} \cdot {\tilde{d}}_{-m}
      - {\tilde{d}}_m \cdot d_m \right)
        + i \theta_0 \sum_{a=1}^4
          \left( s_a^{\dagger} {\tilde{s}}_a^{\dagger}
            - {\tilde{s}}_a  s_a \right).
\EA
The tilde denotes the operators associated with the fictitious system, and these tilde operators satisfy the similar (anti-)commutation relations as the original operators. $\theta$'s are the parameters that will be determined later by the condition that the partial derivatives of free energy with respect to these parameters vanish. For the zero mode of Ramond fermion, we have defined operators $s_a$ and $s_a^{\dagger}$ as
\BA
  s_a^{\dagger}
    &\equiv& \frac{1}{\sqrt{2}} \left( d_0^{2a} + i d_0^{2a+1} \right), \\
  s_a &\equiv& \frac{1}{\sqrt{2}} \left( d_0^{2a} - i d_0^{2a+1} \right),
    \hspace{10mm} a = 1 , \cdots , 4
\EA
These operators satisfy anticommutation relation
\BA
  \left\{ s_a , s_b^{\dagger} \right\} &=& \delta^{ab}, \\
  \left\{ s_a , s_b \right\}
    &=& \left\{ s_a^{\dagger} , s_b^{\dagger} \right\}
      = 0.
\EA
The tilde operators ${\tilde{s}}_a$ and ${\tilde{s}}_a^{\dagger}$ are also defined in the same way, and satisfy the similar anticommutation relations. ${\cal G}_B$, ${\cal G}_{NS}$ and ${\cal G}_R$ are Hermitian operators, namely,
\BA
  {\cal G}_B^{\dagger} = {\cal G}_B, \hspace{10mm}
    {\cal G}_{NS}^{\dagger} = {\cal G}_{NS}, \hspace{10mm}
      {\cal G}_R^{\dagger} = {\cal G}_R.
\EA
Using these generators, the thermal vacua are defined as
\BA
  \left| 0_{1NS} \left( \theta \right) \right\rangle
    &\equiv& e^{- i G_{1NS}} | 0 \rangle \rangle, \\
  \left| 0_{1R} \left( \theta \right) \right\rangle
    &\equiv& e^{- i G_{1R}} | 0 \rangle \rangle,
\EA
where the vacuum state $| 0 \rangle \rangle$ represents the vacuum states in the doubled Hilbert space, and satisfies
\BA
  \alpha_l | 0 \rangle \rangle
    = b_r | 0 \rangle \rangle
      = d_m | 0 \rangle \rangle
        = s_a | 0 \rangle \rangle
          = 0,
\EA
for positive $l$, $r$, $m$, and the similar condition for the fictitious operators. Let us decompose into boson, NS fermion and R fermion parts, and calculate each part separately. The boson oscillator part of the vacuum state is calculated as
\BA
  \left| 0_B \left( \theta \right) \right\rangle
    &\equiv& e^{- i {\cal G}_B} | 0 \rangle \rangle
      \nonumber \\
  &=& \exp \left[ \sum_{l=1}^{\infty} \frac{1}{l} \ \theta_l
    \left( \alpha_{-l} \cdot {\tilde{\alpha}}_{-l}
      - {\tilde{\alpha}}_l \cdot \alpha_l \right) \right]
        | 0 \rangle \rangle
          \nonumber \\
  &=& \prod_{l=1}^{\infty} \left\{
    \left( \frac{1}{\cosh (\theta_l)} \right)^8
      \exp \left[ \frac{1}{l} \ \tanh (\theta_l)
        \alpha_{-l} \cdot {\tilde{\alpha}}_{-l} \right] \right\}
          | 0 \rangle \rangle.
\EA
The fermion parts are computed as
\BA
  \left| 0_{NS} \left( \theta \right) \right\rangle
    &\equiv& e^{- i {\cal G}_{NS}} | 0 \rangle \rangle
      \nonumber \\
  &=& \exp \left[ \sum_{r = \frac{1}{2}}^{\infty} \ \theta_r
    \left( b_{-r} \cdot {\tilde{b}}_{-r}
      - {\tilde{b}}_r \cdot b_r \right) \right]
        | 0 \rangle \rangle
          \nonumber \\
  &=& \prod_{r = \frac{1}{2}}^{\infty}
    \left( \cos (\theta_r) \right)^8
      \exp \left[ \ \tan (\theta_r)
        b_{-r} \cdot {\tilde{b}}_{-r} \right]
          | 0 \rangle \rangle,
\EA
for NS sector, and
\BA
  \left| 0_R \left( \theta \right) \right\rangle
    &\equiv& e^{- i {\cal G}_R} | 0 \rangle \rangle
      \nonumber \\
  &=& \exp \left[ \sum_{m=1}^{\infty} \ \theta_m
    \left( d_{-m} \cdot {\tilde{d}}_{-m}
      - {\tilde{d}}_m \cdot d_m \right)
        + \theta_0 \sum_{a=1}^4
          \left( s_a^{\dagger} {\tilde{s}}_a^{\dagger}
            - {\tilde{s}}_a  s_a \right) \right]
              | 0 \rangle \rangle
                \nonumber \\
  &=& \prod_{m=1}^{\infty}
    \left( \cos (\theta_m) \right)^8
      \exp \left[ \ \tan (\theta_m)
        d_{-m} \cdot {\tilde{d}}_{-m} \right]
          \prod_{a=1}^4 \cos (\theta_0)
            \exp \left[ \ \tan (\theta_0)
              s_a^{\dagger} {\tilde{s}}_a^{\dagger} \right]
                | 0 \rangle \rangle,
                  \nonumber \\
\EA
for R sector. Using these states, the thermal vacuum state for a single string can be expressed as
\BA
  \left| 0_{1NS} (\theta) \right\rangle
    &\equiv& e^{- i G_{1NS}} | 0 \rangle \rangle
      \left| k^+ \right\rangle
        \left| \mbox{\boldmath $k$} \right\rangle
          \nonumber \\
  &=& \left| 0_B \left( \theta \right) \right\rangle
    \left| 0_{NS} \left( \theta \right) \right\rangle
      \left| k^+ \right\rangle
        \left| \mbox{\boldmath $k$} \right\rangle
          \nonumber \\
  &=& \prod_{l=1}^{\infty} \left\{
    \left( \frac{1}{\cosh (\theta_l)} \right)^8
      \exp \left[ \frac{1}{l} \ \tanh (\theta_l)
        \alpha_{-l} \cdot {\tilde{\alpha}}_{-l} \right] \right\}
          \nonumber \\
  && \hspace{10mm}
    \times \prod_{r = \frac{1}{2}}^{\infty}
      \left\{ \left( \cos (\theta_r) \right)^8
        \exp \left[ \ \tan (\theta_r)
          b_{-r} \cdot {\tilde{b}}_{-r} \right] \right\}
            | 0 \rangle \rangle
              \left| k^+ \right\rangle
                \left| \mbox{\boldmath $k$} \right\rangle,
\EA
for NS string and
\BA
  \left| 0_{1R} (\theta) \right\rangle
    &\equiv& e^{- i G_{1R}} | 0 \rangle \rangle
      \left| k^+ \right\rangle
        \left| \mbox{\boldmath $k$} \right\rangle
          \nonumber \\
  &=& \left| 0_B \left( \theta \right) \right\rangle
    \left| 0_R \left( \theta \right) \right\rangle
      \left| k^+ \right\rangle
        \left| \mbox{\boldmath $k$} \right\rangle
          \nonumber \\
  &=& \prod_{l=1}^{\infty} \left\{
    \left( \frac{1}{\cosh (\theta_l)} \right)^8
      \exp \left[ \frac{1}{l} \ \tanh (\theta_l)
        \alpha_{-l} \cdot {\tilde{\alpha}}_{-l} \right] \right\}
          \nonumber \\
  && \hspace{10mm}
    \times \prod_{m=1}^{\infty}
      \left\{ \left( \cos (\theta_m) \right)^8
        \exp \left[ \ \tan (\theta_m)
          d_{-m} \cdot {\tilde{d}}_{-m} \right] \right\}
            \nonumber \\
  && \hspace{20mm}
    \times \prod_{a=1}^4 \cos (\theta_0)
      \exp \left[ \ \tan (\theta_0)
        s_a^{\dagger} {\tilde{s}}_a^{\dagger} \right]
          | 0 \rangle \rangle
            \left| k^+ \right\rangle
              \left| \mbox{\boldmath $k$} \right\rangle,
\EA
for R string. We can compute the expectation values for thermodynamical quantities for a single string by using these thermal vacuum states.

\section{Free Energy and Partition Function for a Single String}
\label{sec:FZsingle}

In this section we compute the free energies and the partition functions for a single string. Let us first consider the free energies for a single string. Following the recipe for TFD, we can compute them from the thermal vacuum states, Hamiltonian operators and appropriate entropy operators. They can be expressed as
\BA
  F_{1NS,R} (\theta)
    = \left\langle 0_{1NS,R} (\theta) \left|
      \left( H_{1NS,R} - \ \frac{1}{\beta} \ K_{1NS,R} \right)
        \right| 0_{1NS,R} (\theta) \right\rangle,
\label{eq:1free}
\EA
From (\ref{eq:Z1LC}), the Hamiltonian operators for a single string can be expressed as
\BA
  H_{1NS,R} = \frac{1}{\sqrt{2}}
    \left( k^+ + \frac{| \mbox{\boldmath $k$} |^2 + M_{NS,R}^2}{2 k^+} \right).
\EA
The entropy operators for a single string are defined as
\BA
  K_{1NS} &\equiv& - \sum_{l=1}^{\infty} \frac{1}{l}
    \left\{ \alpha_{-l} \cdot \alpha_l
      \ln \sinh^2 \theta_l
        - \alpha_l \cdot \alpha_{-l}
          \ln \cosh^2 \theta_l \right\}
            \nonumber \\
  && \hspace{10mm}
    - \sum_{r = \frac{1}{2}}^{\infty}
      \left\{ b_{-r} \cdot b_r \ln \sin^2 \theta_r
        + b_r \cdot b_{-r} \ln \cos^2 \theta_r \right\},
\EA
for NS string, and
\BA
  K_{1R} &\equiv& - \sum_{l=1}^{\infty} \frac{1}{l}
    \left\{ \alpha_{-l} \cdot \alpha_l
      \ln \sinh^2 \theta_l
        - \alpha_l \cdot \alpha_{-l}
          \ln \cosh^2 \theta_l \right\}
            \nonumber \\
  && \hspace{10mm}
    - \sum_{m=1}^{\infty} \left\{ d_{-m} \cdot d_m \ln \sin^2 \theta_m
      + d_m \cdot d_{-m} \ln \cos^2 \theta_m \right\}
        \nonumber \\
  && \hspace{20mm}
    - \sum_{a=1}^4 \left\{ s_a^{\dagger} s_a \ln \sin^2 \theta_0
      + s_a s_a^{\dagger} \ln \cos^2 \theta_0 \right\},
\EA
for R string. Let us decompose into boson, NS fermion and R fermion parts, and calculate each part separately. The boson oscillator part can be calculated as
\BA
  f_B \left( \theta \right)
    &\equiv& \left\langle 0_B \left( \theta \right) \left|
      \left[ \frac{1}{2 \sqrt{2} \ \alpha' k^+}
        \sum_{l=1}^{\infty} \alpha_{-l} \cdot \alpha_l
          \right. \right. \right. \nonumber \\
  && \hspace{20mm} \left. \left. \left.
    + \sum_{l=1}^{\infty} \frac{1}{\beta l}
      \left\{ \alpha_{-l} \cdot \alpha_l
        \ln \sinh^2 \theta_l
          - \alpha_l \cdot \alpha_{-l}
            \ln \cosh^2 \theta_l \right\} \right]
              \right| 0_B \left( \theta \right) \right\rangle
                \nonumber \\
  &=& \sum_{l=1}^{\infty}
    \left\{ - \ \frac{8}{\beta} \ \ln \cosh^2 \theta_l
      + 8l \sinh^2 \theta_l \left( \frac{1}{2 \sqrt{2} \ \alpha' k^+}
        + \frac{1}{\beta l} \ \ln \tanh^2 \theta_l \right) \right\}.
\EA
NS fermion part and R fermion one are given by
\BA
  f_{NS} \left( \theta \right)
    &\equiv& \left\langle 0_{NS} \left( \theta \right) \left|
      \left[ \frac{1}{2 \sqrt{2} \ \alpha' k^+}
        \sum_{r = \frac{1}{2}}^{\infty}
          \left( r b_{-r} \cdot b_r - \ \frac{1}{2} \right)
            \right. \right. \right. \nonumber \\
  && \hspace{20mm} \left. \left. \left.
    + \frac{1}{\beta} \sum_{r = \frac{1}{2}}^{\infty}
      \left\{ b_{-r} \cdot b_r \ln \sin^2 \theta_r
        + b_r \cdot b_{-r} \ln \cos^2 \theta_r \right\} \right]
          \right| 0_{NS} \left( \theta \right) \right\rangle
            \nonumber \\
  &=& - \ \frac{1}{4 \sqrt{2} \ \alpha' k^+}
    \nonumber \\
  && \hspace{10mm}
    + \sum_{r = \frac{1}{2}}^{\infty}
      \left\{ \frac{8}{\beta} \ \ln \cos^2 \theta_r
        + 8 \sin^2 \theta_r \left( \frac{r}{2 \sqrt{2} \ \alpha' k^+}
          + \frac{1}{\beta} \ \ln \tan^2 \theta_r \right) \right\},
\EA
and
\BA
  f_R \left( \theta \right)
    &\equiv& \left\langle 0_R \left( \theta \right) \left|
      \left[ \frac{1}{2 \sqrt{2} \ \alpha' k^+}
        \sum_{m=1}^{\infty} m d_{-m} \cdot d_m
          \right. \right. \right. \nonumber \\
  && \hspace{20mm}
    + \frac{1}{\beta} \sum_{m=1}^{\infty}
      \left\{ d_{-m} \cdot d_m \ln \sin^2 \theta_m
        + d_m \cdot d_{-m} \ln \cos^2 \theta_m \right\}
          \nonumber \\
  && \hspace{40mm} \left. \left. \left.
    + \frac{1}{\beta} \sum_{a=1}^4
      \left\{ s_a^{\dagger} s_a \ln \sin^2 \theta_0
        + s_a s_a^{\dagger} \ln \cos^2 \theta_0 \right\} \right]
          \right| 0_R \left( \theta \right) \right\rangle
            \nonumber \\
  &=& \sum_{m=1}^{\infty}
    \left\{ \frac{8}{\beta} \ \ln \cos^2 \theta_m
      + 8 \sin^2 \theta_m \left( \frac{m}{2 \sqrt{2} \ \alpha' k^+}
        + \frac{1}{\beta} \ \ln \tan^2 \theta_m \right) \right\}
          \nonumber \\
  && \hspace{40mm}
    + \frac{4}{\beta}
      \left( \ln \cos^2 \theta_0
        + \sin^2 \theta_0 \ln \tan^2 \theta_0 \right),
\EA
respectively. Collecting all these together, we can express the free energies as
\BA
  F_{1NS} (\theta)
    &=& \frac{1}{\sqrt{2}}
      \left( k^+ + \frac{| \mbox{\boldmath $k$} |^2}{2 k^+} \right)
        + \frac{|T|^2}{\sqrt{2} \ \alpha' k^+}
          + f_B \left( \theta \right) + f_{NS} \left( \theta \right), \\
  F_{1R} (\theta)
    &=& \frac{1}{\sqrt{2}}
      \left( k^+ + \frac{| \mbox{\boldmath $k$} |^2}{2 k^+} \right)
        + \frac{|T|^2}{\sqrt{2} \ \alpha' k^+}
          + f_B \left( \theta \right) + f_R \left( \theta \right).
\EA
We here determine the $\theta$ parameters from the condition that the derivatives of these free energies with respect to the $\theta$ parameters vanish as we have mentioned before. Since only the boson oscillator part $f_B$ of $F_{1NS}$ and $F_{1R}$ depends on $\theta_l$, the above condition means that
\BA
  \frac{\partial}{\partial \theta_l} \ f_B \left( \theta \right)
    = 16 \sinh \theta_l \cosh \theta_l
      \left( \frac{l}{2 \sqrt{2} \ \alpha' k^+}
        + \frac{1}{\beta} \ \ln \tanh^2 \theta_l \right)
          = 0.
\EA
If we suppose that $\theta_l > 0$, then we obtain
\BA
  \tanh \theta_l
    = \exp \left( - \ \frac{\beta l}{4 \sqrt{2} \ \alpha' k^+} \right).
\EA
Similarly, from the derivatives of free energies with respect to $\theta_r$, $\theta_m$ and $\theta_0$, we obtain
\BA
  \tan \theta_r
    &=& \exp \left( - \ \frac{\beta r}{4 \sqrt{2} \ \alpha' k^+}
      \right), \\
  \tan \theta_m
    &=& \exp \left( - \ \frac{\beta m}{4 \sqrt{2} \ \alpha' k^+}
      \right), \\
  \tan \theta_0 &=& 1.
\EA
$\theta_0$ can be included in the $m=0$ case of $\theta_m$. If we restrict $0 < \theta_0 < \pi / 2$, then we obtain $\theta_0 = \pi /4$. From these relations between $\theta$ and $\beta$, we can represent each part of the free energies as functions of $\beta$ as
\BA
  f_B \left( \beta \right)
    &=& \frac{8}{\beta} \sum_{l=1}^{\infty}
      \ln \left[ 1
        - \exp \left( - \ \frac{\beta l}{2 \sqrt{2} \ \alpha' k^+}
          \right) \right], \\
  f_{NS} \left( \beta \right)
    &=& - \ \frac{1}{4 \sqrt{2} \ \alpha' k^+}
      - \ \frac{8}{\beta} \sum_{r = \frac{1}{2}}^{\infty}
        \ln \left[ 1
          + \exp \left( - \ \frac{\beta r}{2 \sqrt{2} \ \alpha' k^+}
            \right) \right], \\
  f_R \left( \beta \right)
    &=& - \ \frac{8}{\beta} \sum_{m=1}^{\infty}
      \ln \left[ 1
        + \exp \left( - \ \frac{\beta m}{2 \sqrt{2} \ \alpha' k^+}
          \right) \right]
            - \ \frac{4}{\beta} \ \ln 2.
\EA
Gathering contribution from all the parts, we obtain the free energies as
\BA
  F_{1NS} (\beta)
    &=& \frac{1}{\sqrt{2}}
      \left( k^+ + \frac{| \mbox{\boldmath $k$} |^2}{2 k^+} \right)
        + \frac{|T|^2}{\sqrt{2} \ \alpha' k^+}
          + \frac{8}{\beta} \sum_{l=1}^{\infty}
            \ln \left[ 1
              - \exp \left( - \ \frac{\beta l}
                {2 \sqrt{2} \ \alpha' k^+} \right) \right]
                  \nonumber \\
  && \hspace{20mm}
    - \ \frac{1}{4 \sqrt{2} \ \alpha' k^+}
      - \ \frac{8}{\beta} \sum_{r = \frac{1}{2}}^{\infty}
        \ln \left[ 1
          + \exp \left( - \ \frac{\beta r}{2 \sqrt{2} \ \alpha' k^+}
            \right) \right],
\label{eq:F1b}
\EA
for a single NS string, and
\BA
  F_{1R} (\beta)
    &=& \frac{1}{\sqrt{2}}
      \left( k^+ + \frac{| \mbox{\boldmath $k$} |^2}{2 k^+} \right)
        + \frac{|T|^2}{\sqrt{2} \ \alpha' k^+}
          + \frac{8}{\beta} \sum_{l=1}^{\infty}
            \ln \left[ 1
              - \exp \left( - \ \frac{\beta l}
                {2 \sqrt{2} \alpha' k^+} \right) \right]
                  \nonumber \\
  && \hspace{30mm}
    - \ \frac{8}{\beta} \sum_{m=1}^{\infty}
      \ln \left[ 1
        + \exp \left( - \ \frac{\beta m}{2 \sqrt{2} \ \alpha' k^+}
          \right) \right]
            - \ \frac{4}{\beta} \ \ln 2,
\label{eq:F1f}
\EA
for a single R string. These are the free energies we can derive directly from the thermal vacuum state for a single string.

Now we have finished computing the free energies as the functions of $\beta$. However, these are the free energies only for a single string, and are not useful for analysis of the thermodynamical system of strings and branes. We need to derive thermodynamic potential for multiple strings. For this purpose, we compute the partition functions for a single string in the remaining part of this section, and derive the free energy for multiple strings from them in the next section. We can compute the partition functions for a single string from the following equation:
\BA
  Z_{1NS,R} (\beta)
    = \frac{\sqrt{2} \ v_p}{(2 \pi)^p}
      \int_0^{\infty} d k^+ \int_{- \infty}^{\infty} d^{p-1} k \ 
        \exp \left( - \beta F_{1NS,R} \right),
\label{eq:Z1def}
\EA
where $v_p$ is the volume of the D$p$-$\overline{\textrm{D$p$}}$ pair which we are considering. The overall factor $\sqrt{2}$ comes from the relation between $k^+$ and $p^1$ in (\ref{eq:p+}). It is noteworthy that we should take integral over momentum to derive these partition functions. Following the idea of TFD, we can obtain thermodynamical variables by only computing the expectation value of the corresponding operators in the thermal vacuum state. However, we are forced to perform such a calculation, because we are treating energy related to momentum in the single string Hamiltonian as the shift of energy. We expect that this problem should be resolved when we use second quantized string field theory to compute thermodynamical variables in TFD. Substituting (\ref{eq:F1b}) into (\ref{eq:Z1def}), we obtain
\BA
  Z_{1NS} (\beta)
    &=& \frac{\sqrt{2} \ v_p}{(2 \pi)^p}
      \int_0^{\infty} d k^+ \ 
        \left( \frac{2 \sqrt{2} \ \pi k^+}{\beta}
          \right)^{\frac{p-1}{2}} \ 
            \exp \left( - \ \frac{\beta k^+}{\sqrt{2}}
              - \ \frac{\beta |T|^2}{\sqrt{2} \ \alpha' k^+}
                + \frac{\beta}{4 \sqrt{2} \ \alpha' k^+} \right)
                  \nonumber \\
  && \hspace{60mm}
    \times \frac{{\displaystyle \prod_{r = \frac{1}{2}}^{\infty}}
      \left\{ 1
        + \exp \left( - \ \frac{\beta r}{2 \sqrt{2} \ \alpha' k^+}
          \right) \right\}^8}
            {{\displaystyle \prod_{l=1}^{\infty}} \left\{ 1
              - \exp \left( - \ \frac{\beta l}{2 \sqrt{2} \ \alpha' k^+}
                \right) \right\}^8}.
\EA
Let us change integration variable to
\BA
  \tau \equiv \frac{\sqrt{2} \ \pi \beta}{{\beta_H}^2 k^+}
    = \frac{\beta}{4 \sqrt{2} \ \pi \alpha' k^+},
\EA
where $\beta_H$ represents the inverse of the Hagedorn temperature, and is defined as
\BA
  \beta_H = 2 \pi \sqrt{2 \alpha'}.
\EA
The partition function for a single NS string on a D$p$-$\overline{\textrm{D$p$}}$ pair is given by
\BA
  Z_{1NS} (\beta)
    = \frac{(2 \pi)^4 \beta v_p}{{\beta_H}^{p+1}}
      \int_0^{\infty} \frac{d \tau}{\tau} \ \tau^{- \frac{p+1}{2}}
        e^{- 4 \pi |T|^2}
          \left\{ \frac{\vartheta_3 (0 | i \tau)}
            {{\vartheta_1}' (0 | i \tau)} \right\}^4
              \exp \left( - \ \frac{\pi \beta^2}{{\beta_H}^2 \tau} \right).
\EA
Similarly, by substituting (\ref{eq:F1f}) into (\ref{eq:Z1def}), we obtain the partition function for a single R string as
\BA
  Z_{1R} (\beta)
    = \frac{(2 \pi)^4 \beta v_p}{{\beta_H}^{p+1}}
      \int_0^{\infty} \frac{d \tau}{\tau} \ \tau^{- \frac{p+1}{2}}
        e^{- 4 \pi |T|^2}
          \left\{ \frac{\vartheta_2 (0 | i \tau)}
            {{\vartheta_1}' (0 | i \tau)} \right\}^4
              \exp \left( - \ \frac{\pi \beta^2}{{\beta_H}^2 \tau} \right).
\EA
It is unsatisfactory to derive the partition function for NS string and that for R string separately. When we succeed to construct the thermal vacuum state for multiple string on the basis of string field theory, we can obtain its free energy by calculating the multiple string version of the expectation value (\ref{eq:1free}). Therefore, it is important to calculate the thermal vacuum state for second quantized strings.

\section{Free Energy for Multiple Strings}
\label{sec:Fmultiple}

From the single string partition functions, we have computed in the previous section, the free energy for multiple strings can be derived from the following formula:
\BA
  F (\beta) = - \sum_{w=1}^{\infty}
    \frac{1}{\beta w} \left\{ Z_{1NS} (\beta w)
      - (-1)^w Z_{1R} (\beta w) \right\}.
\EA
We can easily make sure that this formula holds for ideal gas system of bosons and fermions in general (see, e.g., Ref. \cite{BowickGiddings}). As we have already mentioned, we may regard strings as a collection of particles with mass spectra given by (\ref{eq:MassNS}) and (\ref{eq:MassR}). We should pay attention to the following point. We must impose the GSO projection for the open strings whose two ends attach to the same D$p$-brane or $\overline{\textrm{D$p$}}$-brane, while the opposite GSO projection for the open strings going from a D$p$-brane to a $\overline{\textrm{D$p$}}$-brane or its inverse. Thus, we need to double the result. We finally obtain
\BA
  F (T, \beta)
    &=& - \ \frac{2 (2 \pi)^4 v_p}{\beta_H^{p+1}}
      \int_{0}^{\infty} \frac{d \tau}{\tau} \ 
        \tau^{- \frac{p+1}{2}} e^{-4 \pi |T|^2 \tau}
          \nonumber \\
  && \hspace{20mm}
    \times \left[ \left\{ \frac{\vartheta_3 (0 | i \tau)}
      {{\vartheta_1}' (0 | i \tau)} \right\}^4
        \left\{ \sum_{w=1}^{\infty}
          \exp \left( - \frac{\pi w^2 \beta^2}{\beta_H^2 \tau}
            \right) \right\}
              \right. \nonumber \\
  && \hspace{30mm} \left.
    - \left\{ \frac{\vartheta_2 (0 | i \tau)}
      {{\vartheta_1}' (0 | i \tau)} \right\}^4
        \left\{ \sum_{w=1}^{\infty} (-1)^w
          \exp \left( - \frac{\pi w^2 \beta^2}{\beta_H^2 \tau}
            \right) \right\} \right].
\EA
This equals to the one-loop free energy (\ref{eq:Dp_antiDp_Free}) based on Matsubara formalism. It should be noted that we obtain the same results even if we adopt different method when the both sides of the strip or the cylinder world-sheet are treated on an equal footing. In this sense, our treatment is consistent. Since the free energy is the same as that in the case of Matsubara formalism, the finite temperature effective potential is also the same. Therefore, the thermal brane creation transition occurs near the Hagedorn temperature in the case of a D9-$\overline{\textrm{D9}}$ pair.

\section{Conclusions and Discussions}
\label{sec:conclusion}

We have computed the thermal vacuum states, the free energies and the partition functions for a single string based on TFD. From these partition functions we have derived the free energy for multiple strings. This free energy agrees with that based on Matsubara formalism. This fact reinforces our argument of the choice of the Weyl factors in the case of Matsubara formalism. Both methods lead to the conclusion that the thermal brane creation transition occurs near the Hagedorn temperature in the case of D9-$\overline{\textrm{D9}}$ pairs as long as we are considering the weak coupling region of open strings.

Although we have restricted ourselves to the case of a single D$p$-$\overline{\textrm{D$p$}}$ pair in a flat spacetime, it is expected that we can also obtain free energy in a similar way even in the case of multiple D$p$-$\overline{\textrm{D$p$}}$ pairs and of non-BPS D$p$-branes not only in a flat spacetime but also in a toroidally compactified spacetime. Thus, it is anticipated that the results based the Matsubara formalism are preserved also in these cases.

The form of the thermal vacuum state resembles that of the D-brane boundary state of a closed string if we identify original oscillators with left-moving ones, and fictitious ones with right-moving ones, respectively. In fact, the relation between the D-brane boundary states and the thermal vacuum states in TFD has already been discussed by many authors \cite{DbraneTFD1} \cite{DbraneTFD2} \cite{DbraneTFD3} \cite{DbraneTFD4} \cite{DbraneTFD5}. Then we can also consider vice versa. That is, when we deal with closed superstrings in type II theory in the framework of TFD, we may identify fictitious left-moving oscillators of a closed string with original ones of an open string, and fictitious right-moving oscillators of a closed string with fictitious ones of an open string. For the bosonic closed string case, the thermal vacuum state and the free energy for a single string have been calculated by Abdalla, Gadelha and Nedel \cite{ClosedStringTFD}. It would be interesting that we generalize to the case of the ideal gas of closed superstrings, and compare the thermal vacuum state for closed superstrings to that for open superstrings on D$p$-$\overline{\textrm{D$p$}}$ pairs.

We need to use second quantized string field theory in order to compute the free energy for multiple strings directly from their thermal vacuum state. In general, it is quite difficult to deal with string fields. However, when we consider ideal gas of strings and only concentrate on free strings, we need not consider interactions of strings, and we may obtain free energy for multiple strings without difficulty. We leave this calculation for future work.

Recently much work has been devoted to study black hole firewall, which has advocated by Almheiri, Marolf, Polchinski and Sully \cite{Firewall1} \cite{Firewall2}. It is widely recognized that external observers measure Hawking radiation. They argued that infalling observers encounter high energy quanta, and burn up at the horizon. If we can clarify Hawking radiation in string theory, then we will be able to tackle this problem. In the case of particle theory, Hawking-Unruh effect can be described by TFD \cite{HawkingUnruhTFD}. In order to discuss the black hole firewall from a string theoretical perspective, we need to generalize it to the case of closed strings, which can propagate bulk spacetime. Although it is very difficult to handle string theory in general curved spacetime, TFD has been already applied to closed strings on AdS black hole background \cite{AdSBHStringTFD1} \cite{AdSBHStringTFD2} \cite{AdSBHStringTFD3} \cite{AdSBHStringTFD4} and on pp-wave background \cite{ppwaveStringTFD1} \cite{ppwaveStringTFD2} \cite{ppwaveStringTFD3} \cite{ppwaveStringTFD4}. Although these works are based on first quantized string theory, they will be helpful to the string theoretical approach to the black hole firewall problem.

If it is extremely hot inside of a black hole independent of observers as is suggested by the black hole firewall argument, then spacetime-filling branes such as D9-$\overline{\textrm{D9}}$  pairs are created by the thermal brane creation transition \cite{Hotta4} \cite{Hotta5} \cite{Hotta6} \cite{Hotta8}. This situation is reminiscent of the phase transition in the Plank solid model of Schwarzschild black holes \cite{Hotta3}. We will be able to identify these spacetime-filling branes as the Planck solid if we succeed in showing that effective field theory on these branes is described by some kind of topological field theory \cite{TFT}. Therefore, it would be very interesting to investigate the thermodynamical systems of strings and spacetime-filling branes in curved spacetime in the framework of TFD.

\acknowledgments

The author would like to thank K. Ishikawa for useful discussions and encouragement. He also thanks colleagues at Hokkaido University for useful discussions. He appreciates the Yukawa Institute for Theoretical Physics at Kyoto University. Discussions during the YITP workshop YITP-T-18-04 on "New Frontiers in String Theory 2018" were useful to complete this work.

\end{document}